\documentclass[fleqn,twoside]{article}
\usepackage{espcrc2}

\usepackage{graphicx}

\newcommand{\AmS}{{\protect\the\textfont2
  A\kern-.1667em\lower.5ex\hbox{M}\kern-.125emS}}

\title{Search for clustering of ultra high energy cosmic rays from
 the Pierre Auger Observatory }

\author{Silvia Mollerach, for the Pierre Auger Collaboration
\address[]{Pierre Auger Observatory, av. San Mart\'{\i}n Norte 304, 
(5613)
  Malarg\"ue, Argentina\\
CONICET, Centro At\'omico Bariloche, 8400 Bariloche, Rio Negro, Argentina}%
        \thanks{mollerach@cab.cnea.gov.ar.}}
       
\begin{document}

\begin{abstract}
We present the results of a search for clustering among the highest energy events
detected by the surface detector of the Pierre Auger Observatory
between 1 January 2004 and 31 August 2007. We analyse the
autocorrelation function, in which the number of pairs with angular
separation within an angle  $\alpha$ is compared 
with the expectation from an isotropic distribution. 
Performing a scan in energy above 30\,EeV and in angles 
$\alpha < 30^\circ$  , the most significant excess
of pairs appears  for $E > 57$\,EeV and for a wide range of separation 
angles, $9^\circ < \alpha <  22^\circ$. 
An excess like this has a chance probability of $\sim 2$\% to
arise from an isotropic distribution and appears at the same
energy threshold at which the Pierre Auger Observatory has reported
a correlation of the arrival directions of cosmic rays with nearby
astrophysical objects. 

\vspace{1pc}
\end{abstract}

\maketitle

\section{Introduction}

The identification of the sources of the ultra high energy cosmic rays 
has been one of the main open problems in astrophysics since their discovery. 
The study of their arrival directions is likely to provide significant 
insight into this question.
 
If cosmic rays are charged particles, protons or heavier nuclei, their 
trajectories are expected to be bent by the intervening 
galactic and extragalactic magnetic fields, and their arrival directions 
will not point back to their sources. 
The intensity and orientation of these fields are not well known, but as
the deflections are inversely proportional to the energy, the effect is 
smaller at the largest energies. Thus, it is at the highest energies 
that cosmic ray arrival directions are most likely to trace their sources.

On the other hand, the distance from which ultra high energy protons 
can arrive to the Earth is expected to be limited by the energy losses 
caused by the photo-pion production processes
in the interaction with the cosmic microwave background (Greisen 
Zatsepin Kuzmin, GZK effect 
\cite{GZK}), and similarly nuclei can 
undergo photo-disintegration processes. 
Hence, at energies above $\sim 60$ EeV, 
cosmic rays should mostly come from nearby sources (closer than $\sim 200$ 
Mpc).

These ideas have motivated an extensive search for clustering signals 
at high energies looking for excesses in the number of pairs at different
angular scales.
Small angular scale searches essentially look for cosmic rays coming from
the same source
which would be the expected signal from the strongest 
nearby sources.
At intermediate angular scales, a clustering signal would be an evidence 
of the pattern characterising the distribution of the nearby sources, with 
pairs of events resulting from cosmic rays coming from different sources. 
The angular scale up to which we can expect that pairs of events come from 
the same source depends on the energy of the events, on the (unknown) 
magnitude of intervening magnetic fields, and on the cosmic ray composition. 

Although the data from a number of experiments
have shown a remarkably isotropic distribution of arrival directions, there
has been a claim of small scale clustering at energies larger than 
$40$ EeV by the AGASA experiment \cite{AGASA}. The most recent published
analysis \cite{teshima} reports 8 pairs (five doublets and a triplet) with 
separation smaller than $2.5^\circ$ among the 
59 events with energy above 40 EeV, while 1.7 were expected from an isotropic flux. The probability for this excess to happen by chance 
was estimated to be less than $10^{-4}$. The significance of 
the AGASA clustering result was, however, subject of debate based on the 
concern that the energy threshold and angular separation were not fixed 
a priori. Tinyakov and Tkachev \cite{TT01} computed the penalisation arising
from making a scan in the energy threshold and obtained a probability 
of $3\times 10^{-4}$. Finley and Westerhoff \cite{FW} took also into account 
the penalisation for a scan in the angular scale and obtained a probability
of $3.5\times 10^{-3}$.
The HiRes 
observatory has found no significant clustering signal at any angular scale
up to $5^\circ$ for any energy threshold above $10$ EeV \cite{hires}.
A hint of correlation at scales around $25^\circ$ and energies above 
$40$ EeV, combining data from HiRes stereo, AGASA, Yakutsk and
SUGAR experiments has been pointed out in Refs.~\cite{KS,CMS}.

The Pierre Auger Observatory, whose construction has recently been completed in
Argentina, has been taking data since January 2004. Its integrated exposure, 
up to the end of August 2007, is 9000 km$^2$ yr sr, representing the largest 
one ever attained by an extensive air shower array at ultra high energies. 
Thanks also to its angular and energy resolutions it offers an excellent data 
set for studies of anisotropies in the arrival directions. 

We apply here the autocorrelation technique to the events with 
energy above 30 EeV, scanning over energy threshold and angular separation 
(up to $30^\circ$), with the aim of searching for possible clustering in the 
arrival directions. A preliminary analysis  has been presented 
in Ref.~\cite{ACICRC}.

The most clear anisotropy signal 
of the arrival directions of high energy events (with $E > 57$ EeV) has been 
reported by the Auger collaboration
\cite{agnpaper,longpaper} through a completely independent method by studying
their correlation with the nearby extragalactic matter distribution.
The observed correlation with nearby AGNs from the V\'eron-Cetty and 
V\'eron catalog \cite{VC} was shown to be incompatible with an isotropic 
distribution at more than $99\%$ CL.
The autocorrelation analysis discussed here does not
depend on an {\it a priori} selection of the possible sources location and,
therefore, gives complementary information with respect to that analysis.

\section{The observatory and the data set}
\label{dataset}

The Pierre Auger Southern Observatory is located  in the 
Province of Mendoza, Argentina, at 35.1$^\circ$--35.5$^\circ$ S, 
69.0$^\circ$--69.6$^\circ$ W and 1300--1400 m a.s.l. The growing observatory 
has been in operation since 2004: the data presented here refer to the period 
between 1 January 2004 and 31 August 2007, during which the number of surface 
stations was increasing from 154 to 1388. The surface detector consists of 
10 m$^2 \times$ 1.2 m water Cherenkov tanks spaced by 1500 m, 
covering an area that in the
period of interest ranged from about 200 km$^2$ to around 3000 km$^2$. 
While the experiment has been described in detail elsewhere \cite{augernim}, 
the relevant features with respect to the present analysis will be outlined 
here. 

The trigger requirement for the array is based on a 3-fold coincidence,
that is satisfied when a triangle of neighbouring stations is triggered.  The 
50\% contamination from accidental events is obviated by an appropriate event 
selection \cite{allard1}, whose power for selecting real showers is larger than 
99\%. 

The arrival directions of the showers are obtained through the time of flight 
differences among the triggered stations. 
The angular resolution, defined as the angular 
radius around the true cosmic ray direction that would contain 
$68\%$ of the reconstructed shower directions, is calculated on an event 
by event basis and checked through the correlations with the fluorescence 
detector. It depends on the number of triggered stations and is better than 
2$^\circ$ for 3-fold events ($E <$ 4 EeV), better 
than 1.2$^\circ$ for 4-folds and 5-folds events ($3 < E < 10$ EeV) and better 
than 0.9$^\circ$ for higher multiplicity events ($E >$ 10 EeV) \cite{res2008}. 

The estimator for the primary energy is the reconstructed signal at 1000~m 
from the shower core \cite{spectrum2008}. 
The conversion from this estimator to energy 
is derived experimentally through the use of a subset of showers detected by 
both the surface and the fluorescence detectors \cite{spectrum2008}. The 
energy resolution is 18\%  and the absolute energy scale has a systematic 
uncertainty of 22\% \cite{dawson07}.

In the present analysis, we apply the following cuts to the events:
\begin{itemize}
\item zenith angle $\theta<60^\circ$;
\item core location within the array boundaries: reconstructed core within a 
triangle of active stations and station with the highest signal surrounded by 
at least 5 active tanks;
\item reconstructed energy $E > 30$ EeV.
\end{itemize}
After these cuts there are 203 events above 30 EeV and 81 events above 40 EeV.
For these events the surface detector trigger efficiency is 100\%. 
The acceptance 
is fully saturated and is determined by purely geometrical considerations 
\cite{allard2}, thus allowing an accurate calculation of the exposure even 
with a changing array configuration. The exposure is flat as a function of
$\sin^2\theta$, 
where $\theta$ is the zenith angle, and it is nearly uniform as a function 
of the azimuth angle $\phi$. 
There is a small modulation of the flux in right ascension due
to the dead times and the growth of the array during the data taking period, 
but this effect 
is small, below $1\%$, and can be ignored in the present analysis.

\section{The autocorrelation function analysis}\label{results}
\subsection{The method}

A standard tool for studying anisotropies is the two-point angular 
correlation function. In the following we will use the correlation function given by
the number of pairs separated by less 
than an angle $\alpha$ among the $N$ events with energy larger than a  
given threshold $E$,
\begin{equation}
n_p(\alpha)=\sum_{i=2}^{N} \sum_{j=1}^{i-1} \Theta(\alpha-\alpha_{ij}),
\end{equation}
where $\alpha_{ij}$ is the angular separation between events $i$ and $j$ and
$\Theta$ is the step function.
The expected number of pairs and the $90\%$ CL error bars are obtained by 
generating a large number ($10^6$) of Monte Carlo simulations with the 
same number of events as in the real data set, 
isotropically distributed and modulated by the exposure of the detector. 
The chance probability for any excess of pairs at
a fixed angle $\alpha$ and energy threshold is found from 
the fraction of simulations with a larger or equal number of 
pairs than what is found in the data at the angular 
scale of interest.

The result of the autocorrelation function analysis depends on the
chosen values of $\alpha$ and $E$.  The fact that the deflections
expected from galactic and extragalactic magnetic fields and the
distribution of the sources are largely unknown prevents us from
fixing these values a priori.  The significance 
of an autocorrelation signal at a given angle and energy, when 
these values have not been fixed a priori, is a delicate issue that
has made, for example, the significance of the AGASA small scale clustering 
claim very controversial.
We adopt here the method proposed by Finley and Westerhoff \cite{FW},
in which a scan over the energy threshold and the angular separation 
is performed. 
For each value of $E$ and $\alpha$, we compute the fraction  $f$ of simulations 
having an equal or larger number of pairs than the data 
by generating $10^6$ simulated isotropic data sets modulated by the exposure. 
The fraction $f$ takes here the role of the distance between the
experimental and expected distributions as used in the Kolmogorov-Smirnov test.  
The most relevant clustering 
signal corresponds to the values of $\alpha$ and $E$ that have the smallest 
value of $f$, referred here to as $f_{min}$. 
To establish the statistical significance of this
 deviation, it is necessary to account for the fact that the angular
 bins, as well as the energy ones, are not independent. 
To do that, we perform 10$^5$ isotropic
 simulations with the same number of events as the data and calculate for each
 realisation the most significant deviation $f_{min}^i$. 
 The statistical significance of the deviation from isotropy is the
 integral of the normalised $f_{min}$ distribution above
 $f_{min}^{data}$. 
We can then estimate the probability that such clustering arises by chance 
from an isotropic distribution just from
the fraction of simulations having $f_{min}^i \le f_{min}^{data}$.

\subsection{Results}

To study the autocorrelation function using the scan technique we have to 
fix the range of the parameter space that we explore. We scan up to 
$30^\circ$ in the angular scale, what includes the range where small scale 
signals from point sources and intermediate angular scale signals from 
clustering of the sources are expected to appear. We scan on energies 
above 30 EeV: this range probes the high energy region where anisotropies are
expected in the astrophysical sources scenario, and goes down enough to cover
the energy range where past anisotropy claims with data from previous
experiments using a similar technique were made,
allowing for a possible energy calibration difference of up to
30\% between the experiments. 
 We therefore compute the number of pairs as a function of the separation 
angle from
$1^\circ$ to $30^\circ$ in steps of $1^\circ$, and as a function of the 
number of events starting from the 10 highest energy events and adding 
events one by one in decreasing energy order up to 100 events, and in 
groups of ten up to 200 events, corresponding to a threshold energy of 
30 EeV, and then compare them with isotropic distributed simulations.

\begin{figure}[t]
\includegraphics[angle=0,width=8cm]{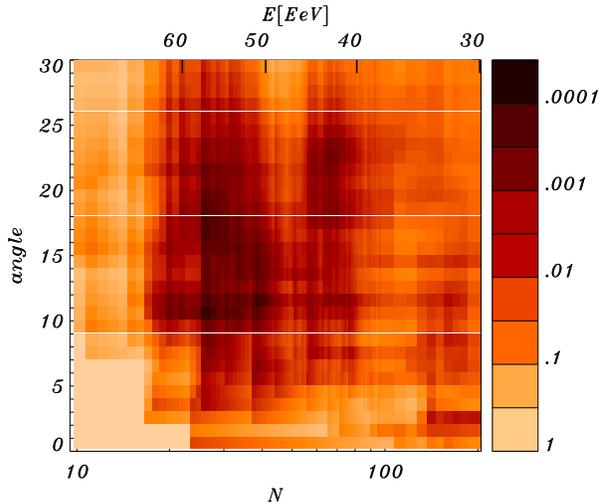}|
\caption{Autocorrelation scan for events with energy above 30 EeV and 
angles up to $30^\circ$. The fraction of simulations with more pairs than
that observed in the data is plotted for each angular scale and number
of events (or energy). }
\label{scan}
\end{figure}

In figure \ref{scan} we present  the fraction of 
simulations with more pairs than the data as a function of the angular 
scale $\alpha$ and the number of events $N$ (or equivalently the 
energy threshold).

A broad region with an excess of pairs appears for energies 
above around 50 EeV, 
at angular scales between about $9^\circ$ to $22^\circ$. In particular, the
minimum value of $f$ ($f_{min} \simeq 1.5 \times 10^{-4}$) 
is found for the
27 highest energy events (corresponding to $E > 57$ EeV), 
for $\alpha = 11^\circ$  (18 observed pairs against 5.2  expected).

\begin{figure}[t]
\includegraphics[angle=270,width=7cm]{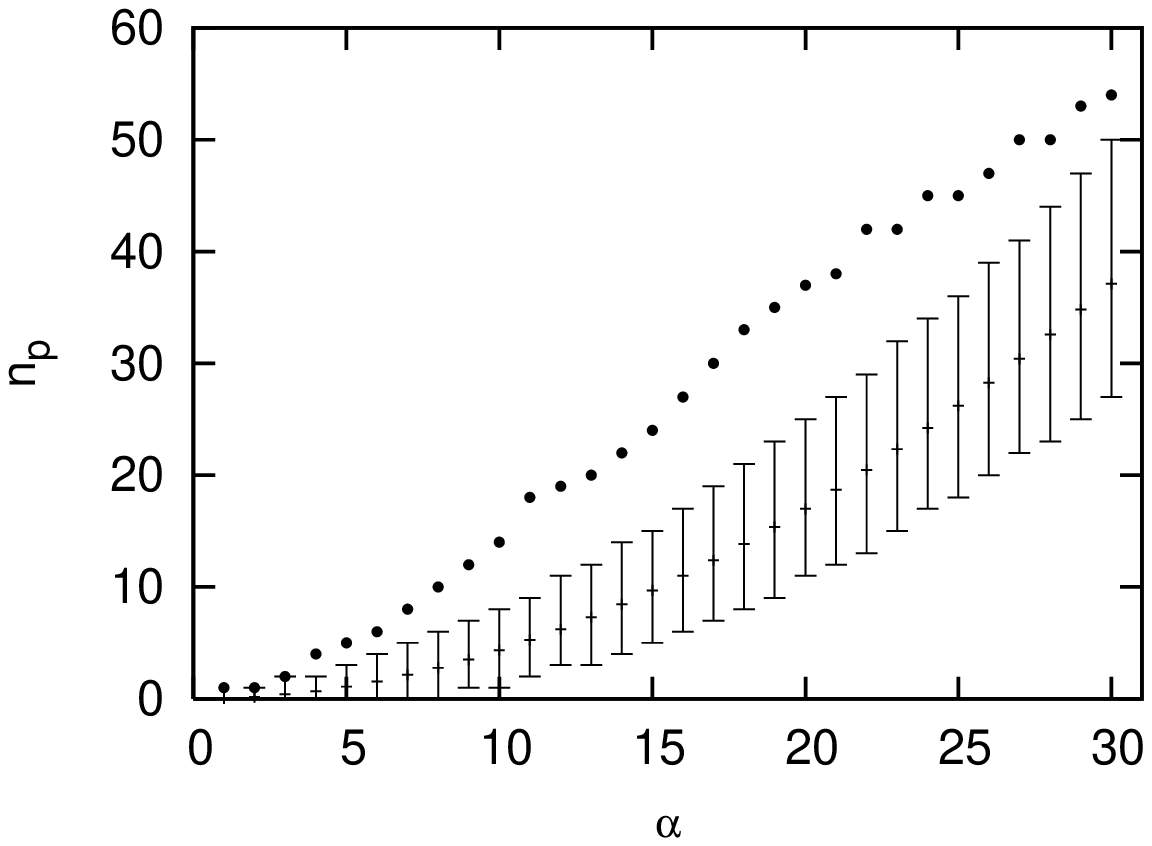}
\includegraphics[angle=270,width=7cm]{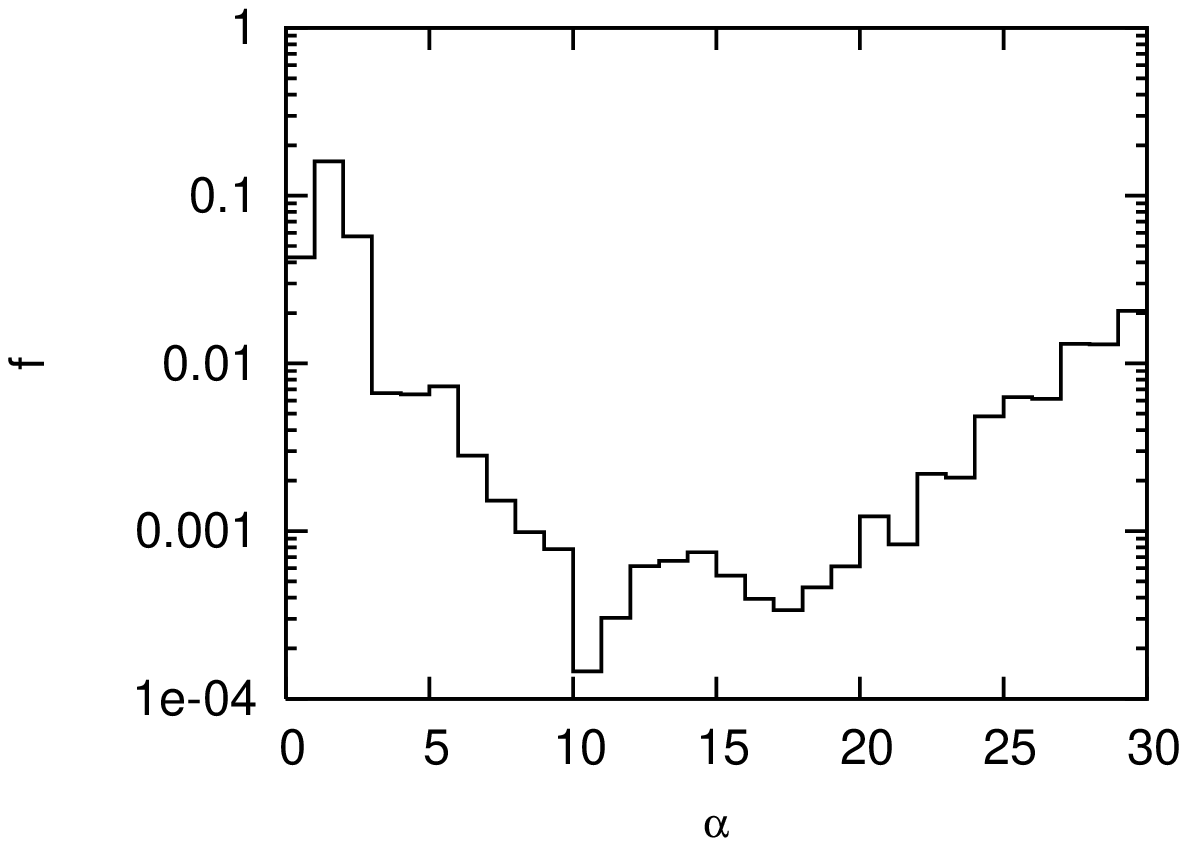}
\caption{
Upper panel: Autocorrelation function for the 27 events with energy larger 
than 57 EeV as a function of the angle (dots) and autocorrelation function 
for an isotropic distribution with $90\%$ confidence level band.
Lower panel: Fraction of isotropic simulations with larger or equal number of 
pairs than the data.
}
\label{cuts}
\end{figure}

In Figure \ref{cuts} the autocorrelation function (upper panel) and 
 the fraction of isotropic simulations with equal or larger number of pairs 
than the data (lower panel)  is shown for these 27 events as a
function of the angular scale. The broad region of low values of $f$ 
($< 10^{-3}$) is visible between 9$^\circ$ and 22$^\circ$, with the
minimum at 11$^\circ$.

The chance probability of a $f_{min} \le 1.5 \times 10^{-4}$ to
arise from an isotropic distribution was estimated by performing the
same scan to $10^5$ simulations, and it is $P \simeq 1.6 \times  10^{-2}$.

\begin{figure}[t]
\includegraphics[width=7cm]{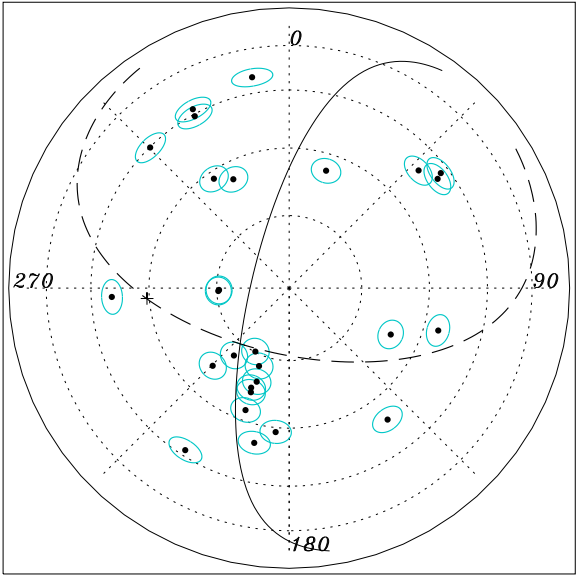}
\caption{
Arrival directions of the 27 events with energy larger than 57 EeV in
equatorial coordinates  centered in the south pole.
The circles around each event have a $5.5^\circ$ radius. The dashed line 
represents the galactic plane (with the galactic center identified by a cross)
and the solid line represents the supergalactic plane.
}
\label{mapa}
\end{figure}

Figure \ref{mapa} shows the arrival directions of the 27 events with 
$E >$ 57 EeV
in equatorial coordinates centered in the south pole. The galactic and 
supergalactic planes are displayed by the dashed and solid lines respectively.
Circles of $5.5^\circ$ around each event guide the eye to identify pairs 
separated by less than $11^\circ$. 
The arrival directions and energies of the
27 events are listed in the Appendix of \cite{longpaper}.

The analysis of the autocorrelation function shows in particular no 
significant clustering signal at the small angular scales
corresponding to the AGASA claim. At energies above 40 EeV, with 81 
detected events, we observe  4 pairs within $2.5^\circ$, 
while 2.5 were expected from an isotropic flux (with a fraction 
$f=0.24$ of isotropic simulations showing a larger or equal number of pairs).
Due to a possible difference in the energy calibration between AGASA and
Auger, the clustering signal reported by AGASA could appear in Auger data
at a lower energy scale.
For energies above 30 EeV the observed number of pairs is 23, while the 
isotropic expectation is 16 ($f = 0.05$). This excess is much smaller than 
the one reported by AGASA.

\section{Conclusions}\label{conclusions}

Using the events recorded by the surface detector of the 
Pierre Auger Observatory between January 2004 and Augist 2007 
we have performed a scan to search for 
possible clustering using the autocorrelation technique. 
The scan has been performed in energy (above 30 EeV, 203 events detected) and 
angular separation (between 1$^\circ$ and 30$^\circ$). 
The most significant excess of pairs is found for $E > 57$ EeV 
(corresponding to 27 events) and 
$\alpha = 11^\circ$,  with a  chance probability of $P = 1.6 \times 10^{-2}$
to arise from an isotropic distribution. 
Above this energy we observe, in fact, a broad region of low probabilities, 
for angular scales between $9^\circ$ and $22^\circ$. At higher energy 
thresholds the chance probabilities become larger, however 
the number of events becomes rather limited.
At lower energies
the autocorrelation function becomes progressively closer to
that expected from an isotropic flux. 

 It should be noted that the energy threshold that maximises the
 significance is the same one as the energy at which the Pierre Auger
 collaboration has observed a departure from isotropy 
 using the correlation of the arrival directions of
 events with nearby AGNs.  It is also the energy at which the cosmic
 ray flux at ultra high energies reported by the Pierre Auger
 collaboration \cite{spectrum2008} is suppressed by 50\% compared to a
  power law extrapolation of the flux measured at lower energies. 
This suppression, if interpreted as due to the
 GZK horizon \cite{GZK}, would imply that nearby sources dominate the
 flux at these energies. As the matter distribution in the nearby
 Universe becomes increasingly anisotropic for smaller distances, it is
 to be expected that the ultra high energy cosmic ray sky should become
 anisotropic at these energies, reflecting the pattern of the 
distribution of the sources.

\end{document}